\documentclass[aps,prl,twocolumn,amsmath,showpacs,superscriptaddress]{revtex4}
\usepackage{bm}
\usepackage{amsmath}
\usepackage{amssymb}
\usepackage{amsfonts}
\usepackage{graphicx}
\usepackage{color}

\begin{document}

\title{Induced Current and Aharonov-Bohm Effect in Graphene}
\author{R. Jackiw}
\affiliation{Department of Physics, Massachusetts Institute of Technology, Cambridge, Massachusetts 02139, USA}

\author{A. I. Milstein}
\affiliation{Budker Institute of Nuclear Physics, 630090
Novosibirsk, Russia}

\author{S.-Y. Pi}
\affiliation{Department of Physics,
Boston University,  Boston, Massachusetts 02215, USA}

\author{I. S. Terekhov}
\affiliation{Budker Institute of
Nuclear Physics, 630090 Novosibirsk, Russia}

\date{\today}

\begin{abstract}
The effect of vacuum polarization in the field of an infinitesimally
thin solenoid at distances much larger than the radius of solenoid
is investigated.  The induced charge density and induced current are
calculated. Though the induced charge density turned out to be zero,
the induced current is finite periodical function of the magnetic
flux $\Phi$. The expression for this function is found exactly in a
value of the flux. The induced current is  equal to zero  at the
integer values of  $\Phi/\Phi_0$ as well as at half-integer values
of this ratio, where $\Phi_0=2\pi\hbar c/e$ is the elementary
magnetic flux. The latter is a consequence of the Furry theorem and
periodicity of the induced current with respect to magnetic flux. As
an example we consider  the graphene in the field of solenoid
perpendicular to the plane of a sample.
\end{abstract}


\maketitle

The Aharonov-Bohm effect \cite{AB59},  scattering of a charged
particle off an infinitesimally thin solenoid, which is absent in
classical electrodynamics,  has been investigated in numerous
papers, see Review \cite{PT89}.  Both non-relativistic \cite{AB59}
and relativistic \cite{AW89,G89, Kh05, Kh08} equations have been
considered. Similar effects having topological origin have been
studied  in quantum field theory in \cite{DJT82,Wilczek89}.
Intensive investigation of the topological effects in
condensed-matter systems has been performed recently   both
experimentally and theoretically in \cite{YPS02,JP07,MT07,JP08}. New
possibilities to study topological effects in  Quantum
Electrodynamics (QED) have appeared after recent successful
fabrication of a monolayer graphite (graphene), see Ref.
\cite{Novoselov1} and recent Review \cite{CN09}.  The single
electron dynamics in  graphene is described by a massless
two-component Dirac equation
\cite{Wallece,McClure,Semenoff84,Gonzalez} so that graphene
represents a peculiar two-dimensional (2D) version of massless QED.
This version is essentially simpler than conventional QED because
effects of retardation are absent in graphene. However, the
 ``fine structure constant''  $\alpha=e^2/\hbar v_F
\sim 1$, since  the Fermi velocity  $v_F \approx 10^6 \mbox{m/s}
\approx c/300$ ($c$ is the velocity of light), and therefore we have
 a strong-coupling version of QED. Below we set $\hbar=c=1$.

Existence of induced charge density in the electric field  of heavy
nucleus due to vacuum polarization is one of the most important
effects of  QED. This problem was investigated in detail in many
papers, see, e.g., Refs. \cite{WichmannKroll,McLerran,Mil2,Zeld}.
Charged impurity screening in graphene can also be treated in terms
of vacuum polarization
\cite{DiVincenzo,nomura,Ando,Hwang,Katsnelson,Shytov,pereira,Subir,Fogler,TMKS07,Kotov08}.
In Ref.\cite{TMKS07}, the induced charge density in graphene has
been investigated analytically using convenient integral
representation for the  Green's function of the two-dimensional
Dirac equation of electron in a Coulomb field.   Calculation of the
induced charge has been performed exactly in the charge of impurity.
In Ref.\cite{TMKS07},   the Green's function has been obtained
following the method based on the operator technique suggested in
Ref.\cite{MS82}. In the present paper, we use similar integral
representation for the Green's function to derive the induced
current in the field of infinitesimally thin solenoid. Calculation
is performed for arbitrary value of the magnetic flux $\Phi$.

 The induced density and induced current in the  vector potential
 $$ \bm A(\bm r)=\frac{\Phi[\bm\nu\times\bm r]}{2\pi r^2}\, ,$$
 where $\bm\nu$ is the unit vector directed along z-axis, have the form
\begin{eqnarray}\label{Inducedrhoj}
\rho_{ind}(\mathbf{r})=-ieN\int_{C}\frac{d\epsilon}{2\pi}\mathrm{Tr}\{
G(\mathbf{r},\mathbf{r}|\epsilon)\}\,, \nonumber\\
 \bm J_{ind}(\mathbf{r})=-ieNv_F\int_{C}\frac{d\epsilon}{2\pi}
\mathrm{Tr}\{\bm\sigma G(\mathbf{r},\mathbf{r}|\epsilon)\}\,,
\end{eqnarray}
where  $N=4$ reflects the spin and valley degeneracies, and the
 Green's function $G({\bm r},{\bm r}'|\epsilon)$
satisfies the equation
\begin{equation}
\label{GFE} \left[\epsilon-v_F\bm\sigma\cdot(\bm p-e\bm A(\bm r))
\right]G({\bm r},{\bm r}'|\epsilon)=\delta(\bm r-{\bm r}')I.
\end{equation}
Here ${\bm{\sigma}}=(\sigma_1,\sigma_2)$, and $\sigma_i$ are the
Pauli matrices; $\mathbf{p}=(p_x,p_y)$ is the momentum operator,
$\bm r=(x,y)$, and $I=\mathrm{diag}\{1,1\}$. The matrixes
${\bm{\sigma}}$ do not act on the spin variables but on the
pseudo-spin ones and the spin degrees of freedom are taken into
account in a factor $N$. According to the Feynman rules, the contour
of integration over $\epsilon$ goes below the real axis in the left
half-plane and above the real axis in the right half-plane of the
complex $\epsilon$ plane. It is convenient to write the function
$G({\bm r},{\bm r}'|\epsilon)$ as
\begin{equation}
\label{GD} G({\bm r},{\bm r}'|\epsilon)
=\left[\epsilon+v_F\bm\sigma\cdot(\bm p-e\bm A) \right]D({\bm
r},{\bm r}'|\epsilon)\, ,
\end{equation}
Where $D({\bm r},{\bm r}'|\epsilon)$ is the Green's function
of the squared Dirac equation,
\begin{eqnarray}
\label{DD} \left[\epsilon^2-v_F^2(\bm p-e\bm A(\bm
r))^2+v_F^2\Phi\delta(\bm r)\sigma_3 \right]D({\bm r},{\bm
r}'|\epsilon)\nonumber\\
=\delta(\bm r-{\bm r}')I.
\end{eqnarray}
For $r\ne 0$ and $r'\ne 0$ , we can omit  the term with
$\delta$-function so that
\begin{equation}
\label{D} \left[\epsilon^2-v_F^2(\bm p-e\bm A(\bm r))^2
\right]D({\bm r},{\bm r}'|\epsilon)=\delta(\bm r-{\bm r}')I.
\end{equation}
The equation (\ref{D}) has regular and singular solutions at $r=0$
and $r'=0$. The Green's function of the Dirac equation with the
magnetic-solenoid field was considered in Ref. \cite{GGS04} taking
into account both regular and singular parts. The singular solutions
originate from the singular behavior at $r=0$ of the vector
potential $\bm A(\bm r)$. Therefore, to find the correct
superposition of regular and singular solutions it is necessary to
perform the appropriate regularization. If we take in mind a real
solenoid, then the natural regularization is the finite radius $R$
of this solenoid, see Ref.\cite{Wilczek89}. Then it is possible to
show that, to calculate the induced charge density and the induced
current at $r\gg R$, we can use the regular Green's function of the
equation (\ref{D}) while the singular part of the Green's function
as well as the term with $\delta$-function in Eq. (\ref{DD})
determine  these quantities at $r\lesssim  R$. The induced current
and the induced charge density at $r\lesssim  R$ depend on the
magnetic field distribution inside the solenoid and, therefore, are
model-dependent. At $r\gg R$, the contribution of the singular part
of the Green's function to the integrand in Eq.(\ref{Inducedrhoj})
contains a factor $(\epsilon R)^\beta$ with some positive $\beta$.
The main contribution to the integral  over $\epsilon$ at $r\gg R$
is given by the region $\epsilon\sim 1/r$ so that the contribution
of the singular part of the Green's function to the induced current
is suppressed by the factor $(R/r)^\beta$. This situation is
completely similar to the problem of a finite nuclear size at the
calculation of the vacuum polarization effects in heavy atoms, see,
e.g., Ref.\cite{LM94}. In quantum field theory it is possible to
consider regularizations different from the solenoid radius $R$. As
a result some uncertainty appears in the prediction of physical
quantities \cite{DJT82,G89,Wilczek89}. In the present paper we
consider the induced charge and induced density at $r\ne 0$ which
are model-independent. Substituting the function $D({\bm r},{\bm
r}'|\epsilon)$ to Eq.(\ref{D}) in the form
\begin{eqnarray}
\label{DA} D({\bm r},{\bm r}'|\epsilon)=
\frac{1}{2\pi}\sum_{m=-\infty}^{\infty} e^{im(\phi-\phi')}
A_m(r,r'|\epsilon) I\,,
\end{eqnarray}
and righthand side of the Eq.(\ref{D}) as
\begin{eqnarray}
\delta({\bm r}-{\bm r}')I=
\frac{\delta(r-r')}{2\pi\sqrt{rr'}}\sum_{m=-\infty}^{\infty}
e^{im(\phi-\phi')}I
\nonumber
\end{eqnarray}
we obtain that the function  $A_m(r,r'|\epsilon)$ satisfies the
equations:
\begin{eqnarray}\label{A}
\left(\frac{\epsilon^2}{v_F^2}+\dfrac{1}{r}\dfrac{\partial}{\partial
r}r\dfrac{\partial}{\partial r}-\dfrac{(m-\gamma)^2}{r^2}
\right)A_m(r,r'|\epsilon)=\dfrac{\delta(r-r')}{v_F^2\sqrt{rr'}}\, ,
\end{eqnarray}
where $\gamma=e\Phi/(2\pi)$. One can see that the function
$A_m(r,r'|\epsilon)$ can be obtained from the Green's function for
the free radial Schr\"odinger equation in the 2D case by the
substitution $m \rightarrow m-\gamma$. A convenient integral
representation for the function  $A_m(r,r'|\epsilon)$ can be
obtained using the operator method developed in Ref.~\cite{MS82} at
the calculation of the Green's function for the Dirac equation of an
electron in a Coulomb field in 3D space. This method   was recently
used in Ref.~\cite{TMKS07} for the case of 2D space.  It follows
from the results of  Ref.~\cite{TMKS07} at zero Coulomb field that
\begin{eqnarray}\label{Am}
&& A_m(r,r'|\epsilon)= -\int_0^{\mu\infty}\dfrac{ds}
  {v_F^2\sinh s}\nonumber\\
&&\times\exp[iE(r+r')\coth s-i\pi\lambda]
J_{2\lambda}\left(\frac{2E\sqrt{rr'}}{\sinh s}\right)\, ,
\end{eqnarray}
where $E=\epsilon/v_F$, $\lambda=|m-\gamma|$,  $J_{2\lambda}(x)$ is
the Bessel function, $\mu=+1$ if $\mbox{Re}\,E>0$ and $\mu=-1$ if
$\mbox{Re}\,E<0$ . The sign  $\mu$ takes into account the analytical
properties of the Green's function.

Taking into account the analytical properties of the Green's
function, the contour of integration with respect to $\epsilon$
 can be deformed to coincide with the imaginary axis.  After these
transformations, we obtain that $\rho_{ind}(\bm r)=0$ as a result of
integration over $\epsilon$. This fact can be easily explained
because, due to the Furry theorem, $\rho_{ind}(\bm{r})$ should be
the odd function of $\gamma=e\Phi/(2\pi)$. However, in this case we
would obtain that $\rho_{ind}(\bm r)$ is pseudoscalar that
contradicts to the parity conservation of the massless 2D Dirac
equation. For $\bm J_{ind}(\mathbf{r})$ we have
\begin{eqnarray}
\label{Inducedj}
 &&\bm J_{ind}(r)=-\frac{eNv_F}{\pi^2
r^2}[\bm\nu\times\bm r]\sum_{m=0}^{\infty}(m-\gamma)\nonumber
\\
&&\times\int\limits_{0}^{\infty}dE\int\limits_{0}^{\infty}
 \frac{ds}{\sinh s}\exp[-2E r\coth s] I_{2\lambda}\left(\frac{2E r}{\sinh s}\right) ,
\end{eqnarray}
where  $I_{2\lambda}(x)$ is the modified Bessel function of the
first kind. We note that $\bm J_{ind}(r)$, Eq.~(\ref{Inducedj}), is
an odd function of $\gamma$, in accordance with the Furry theorem.
To have a possibility to change the order of summation and
integration, we introduce some quantity $\delta\ll 1$ as a lower
limit of integration over $s$. After that we take the integral over
$E$ and obtain
\begin{eqnarray}
\label{Inducedj1}
 && \bm J_{ind}(r)=-\frac{eNv_F}{2\pi^2r^3}[\bm\nu\times\bm
r]\sum_{m=-\infty}^{\infty}(m-\gamma)\nonumber\\
&&\times\int\limits_{\delta}^{\infty}
 \frac{ds}{\sinh s}\,\exp[-2\lambda s] \, .
\end{eqnarray}
As should be,  $\bm J_{ind}(r)$ depends only on the  fractional
 part $\tilde\gamma$ of $\gamma$, $|\tilde\gamma|<1$. The quantity
$\tilde\gamma$ is  $\tilde\gamma=\gamma-n$ for $\gamma>0$ and
$\tilde\gamma=\gamma+n$ for $\gamma<0$, where $n$ is a maximal
integer number less than $|\gamma|$. Then we perform summation over
$m$ and set $\delta=0$. We have
\begin{eqnarray}
\label{Inducedj2}
 &&\bm J_{ind}(r)=\frac{eNv_F}{2\pi^2
r^3}[\bm\nu\times\bm r]\int\limits_{0}^{\infty}
 \frac{ds}{\sinh s}\Bigl[ \tilde\gamma   \exp(-2|\tilde\gamma|s) \nonumber\\
 &&+\tilde\gamma
 \exp(-s)\frac{\cosh(2 \tilde\gamma s)}{\sinh s}
-\frac{\sinh(2\tilde\gamma s)}{2\sinh^2s}\Bigr] \, .
\end{eqnarray}
Taking the integral over $s$ we finally arrive at
\begin{eqnarray}
\label{Inducedj3} \bm J_{ind}(r)=\frac{eNv_F}{16\pi
}F(\tilde\gamma)\,\mbox{curl}\left(\frac{\bm\nu}{r}\right)\,
,\nonumber\\
F(\tilde\gamma)=(1-2|\tilde\gamma|)^2\tan (\pi\tilde\gamma) \, .
\end{eqnarray}
It is interesting that $\bm J_{ind}(r) $ equals to zero at
$|\tilde\gamma|=1/2$. This may be explained as follows. Due to
invariance of $\bm J_{ind}(r)$ under the substitution
$\gamma\rightarrow \gamma-1$  we have
$F(|\tilde\gamma|)=F(|\tilde\gamma|-1)$, and due to the Furry
theorem it should be $F(\tilde\gamma)=-F(-\tilde\gamma)$. From these
two relations  we obtain that $F(\pm 1/2)=0$.

\begin{figure}[h]
\includegraphics[scale=0.85]{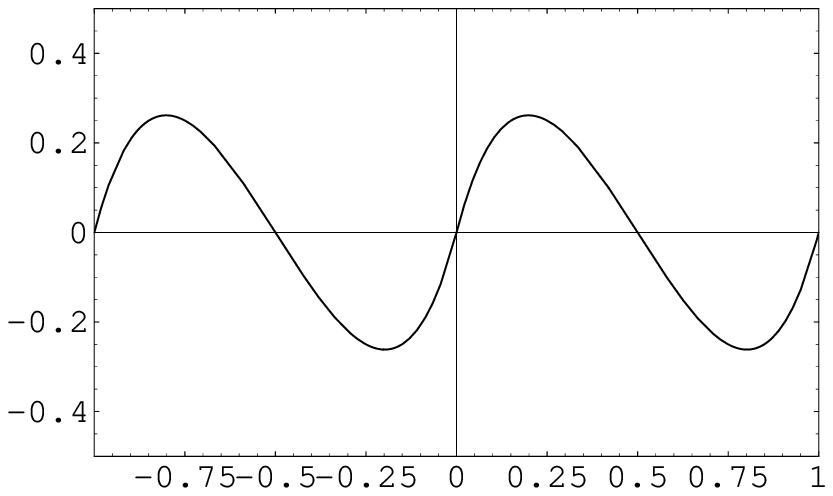}
\begin{picture}(0,0)(0,0)
\put(-102,-8){$\tilde\gamma$}
 \put(-220,68){\rotatebox{90}{$F(\tilde\gamma)$}}
 \end{picture}
\caption{Dependence of the function  $F(\tilde\gamma)$,
Eq.(\ref{Inducedj3}), on the fractional part $\tilde\gamma$ of
$\gamma=e\Phi/(2\pi)$, where $\Phi$ is the magnetic flux.}\label{F}
\end{figure}

The induced charge density and current in the presence of an
infinitesimally thin solenoid were also considered  in Refs.
\cite{S99,S00}. The results of these papers contain the contribution
of both singular and regular parts of the Green's function since a
regularization for the field was not applied. Therefore, the
expressions for the induced charge density and current contain
uncertainty which does not allow one to make any explicit
predictions for these quantities.  Note that our results are in
agreement with the contribution of the regular part of the Green's
function in Ref. \cite{S00} (first term in Eq.(6.14)).

To summarize, we have investigated the effect of vacuum polarization
in the field of an infinitesimally thin solenoid at distances much
larger than the radius of solenoid.  It  turns out that the induced
charge density is zero. We have derived exactly in a magnetic flux
the  expression for the induced current. This current is a periodic
function of the magnetic flux and is equal to zero not only at the
integer values of  $\Phi/\Phi_0$ but also at half-integer values of
this ratio. Though the system considered in our paper  consists of
graphene and a solenoid perpendicular to the plane of a sample, the
results can be easily  generalized for another systems such as
studied in Ref.\cite{YPS02}.

A.I.M. and I.S.T. are very grateful to R.N.~Lee  for valuable
discussions, while R.J. benefitted from conversations with
V.N.~Kotov, Z.~Tesanovic and S.~Sachdev.   A.I.M. gratefully
acknowledges the Max-Planck- Institute for Nuclear Physics,
Heidelberg, for the warm hospitality and financial support during
his visit. The work was supported in part by DOE grants
DE-FG02-05ER41360, DE-FG02-91ER40676 and RFBR
 grants  08-02-91969 and 09-02-00024.

\end{document}